\date{}
\documentclass[showpacs]{revtex4}
\usepackage{graphicx,psfrag,amsmath,amssymb,amsfonts,bbm,latexsym,color,dcolumn}
\newcommand{\be}{\begin{equation}}
\newcommand{\bea}{\begin{eqnarray}}
\newcommand{\ee}{\end{equation}}
\newcommand{\eea}{\end{eqnarray}}

\newcommand{\mn}{{\mu\nu}}
\newcommand{\tmn}{\langle T_\mn\rangle}

\begin{document}
\title{Conformal invariance and apparent universality of semiclassical gravity}
\author{A. Garbarz }
\author{G. Giribet }
\author{F. D. Mazzitelli }
\affiliation{Departamento de F\'\i sica {\it Juan Jos\'e
Giambiagi}, FCEyN UBA, Facultad de Ciencias Exactas y Naturales,
Ciudad Universitaria, Pabell\' on I, 1428 Buenos Aires,
Argentina.}
\date{today}

\begin{abstract} In a recent work,
it has been pointed out that certain observables of the massless
scalar field theory in a static spherically symmetric background
exhibit a universal behavior  at large distances. More precisely, it
was shown that, unlike what happens in the case the coupling to the
curvature $\xi $ is generic, for the special cases $\xi=0$ and $\xi
= 1/6$ the large distance behavior of the expectation value $\left<
{T^{\mu }}_{\nu } \right> $ turns out to be independent of the
internal structure of the gravitational source. Here, we address a
higher dimensional generalization of this result: We first compute
the difference between a black hole and a static spherically
symmetric star for the observables $\left< \phi ^2 \right> $ and
$\left< {T^{\mu }}_{\nu } \right> $ in the far field limit.Thus, we
show that the conformally invariant massless scalar field theory in
a static spherically symmetric background exhibits such universality
phenomenon in $D\geq 4$ dimensions. Also, using the one-loop
effective action, we compute $\left< {T^{\mu }}_{\nu } \right> $ for
a weakly gravitating object. These results lead to the explicit
expression of the expectation value $\left< {T^{\mu }}_{\nu }
\right> $ for a Schwarzschild-Tangherlini black hole in the far
field limit. As an application, we obtain quantum corrections to the
gravitational potential in $D$ dimensions, which for $D=4$ are shown
to agree with the one-loop correction to the graviton propagator
previously found in the literature.

\end{abstract}


\maketitle






\section{Introduction}

In quantum field theory in curved spaces, vacuum polarization
effects exhibit, in general, a non-local dependence on the spacetime
metric. For example, particle production in Robertson Walker metrics
depend on the whole evolution of the scale factor \cite{BD}. More
closely to the present work, in static and spherically symmetric
geometries, the expectation value of the energy momentum tensor
evaluated outside a weakly gravitating object depends on its inner
structure \cite{satz}. More generally, for arbitrary metrics, a
covariant expansion of the effective action in powers of the
curvature tensor is explicitly non-local \cite{Avra,bv}.

In a recent work,  Anderson and Fabbri \cite{Anderson}  studied what
they called ``apparent universality in semiclassical gravity'',
which is exhibited by certain observables corresponding to the
theory of a massless quantum scalar field on static spherically
symmetric backgrounds.  More specifically, they have shown that, far
from the classical gravitational source, the mean value
$\left\langle \phi ^{2}\right\rangle $ in the Boulware state, does
not depend on the internal structure of the source when the scalar
field is minimally coupled to the curvature, i.e. the result is the
same for a black hole, a neutron star, or a weakly gravitational
object, as long as they are static and spherically symmetric. The
situation for $\left\langle T_{\mu \nu }\right\rangle $ is
different, because the universal behavior holds both for minimal and
conformal couplings.

In this paper, we will be concerned with the computation of the
expectation values $\left\langle \phi ^{2}\right\rangle $ and
$\left\langle T_{\mu \nu }\right\rangle $, corresponding to a
massless scalar field $\phi$ formulated on a $D$-dimensional
spherically symmetric background, being such expectation values
defined with respect to the Boulware state. In $D$ dimensions, and
in the large distance limit, these observables are typically given
by
\begin{equation}
\left\langle \phi ^{2}(x)\right\rangle \simeq \frac{aM}{r^{2D-5}},\qquad
\left\langle T_{\mu \nu }(x)\right\rangle \simeq \frac{bM}{r^{2D-3}},
\label{Eq1}
\end{equation}%
where $M$ is the mass of the gravitational background, while $a$ and
$b$ are two numerical coefficients that depend on $D$, the coupling
$\xi$, and may also depend on the internal structure of the
gravitational source.

In the case the gravitational object is a star \footnote{We shall
call \textit{Star} to any static spherically symmetric distribution
of matter without a horizon, although we will sometimes fall in
redundancies like ``spherically symmetric star'' just to emphasize
the importance of the symmetry.}, these coefficients are obtained by
reading the large distance behavior of non-local terms arising in
the one-loop computation. On the other hand, in the case of a black
hole, these coefficients may be obtained by using the method of
\cite{ASH,christensen}. In fact, in a generic case, the precise
values of $a$ and $b$ do depend on whether a horizon exists or not.
Nevertheless, as it was pointed out by Anderson and Fabbri in Ref.
\cite{Anderson}, there exist very special cases where (\ref{Eq1})
exhibit some kind of universality, so that the large distance limit
of the expectation values turn out to be independent on the nature
of the gravitational object. Here, we will study this universality
phenomenon, which can be seen to occur in the minimally coupled and
conformally coupled scalar field theories.

In \cite{Anderson}, it was shown that in the four-dimensional conformally
coupled case ($\xi =1/6$ with $D=4$) the large distance behavior of $%
\left\langle T_{\mu \nu }\right\rangle $ results independent on
whether the gravitational object is either a black hole or a star.
This also occurs for the minimally coupled case ($\xi =0$), for both
$\left\langle T_{\mu \nu }\right\rangle $ and $\left\langle \phi
^{2}\right\rangle $. We can express these agreements by saying that
in the large distance limit it happens that
\begin{equation}
\Delta \left\langle T_{\mu \nu }(x)\right\rangle =\left\langle T_{\mu \nu
}(x)\right\rangle _{\text{Star}}-\left\langle T_{\mu \nu }(x)\right\rangle _{%
\text{BH}}\sim \frac{\xi (\xi -1/6)M}{r^{5}}\text{ }+\mathcal{O}%
(M^{2}/r^{6}),  \label{A}
\end{equation}%
and
\begin{equation}
\Delta \langle \phi ^{2}(x)\rangle =\langle \phi ^{2}(x)\rangle _{\text{Star}%
}-\langle \phi ^{2}(x)\rangle _{\text{BH}}\sim \frac{\xi M}{r^{3}}\text{ }+%
\mathcal{O}(M^{2}/r^{4})  \label{B}
\end{equation}

As already pointed out in \cite{Anderson}, the coincidence of the
results for minimal coupling can be traced back to the fact that
the large distance behavior of the observables is determined by
the $s$-wave in the low frequency limit. The field modes turn out
to be independent of the metric in this limit, so the differences
$\Delta \left\langle T_{\mu \nu }\right\rangle $ and
$\Delta\left\langle \phi ^{2}\right\rangle $ vanish.

In the absence of a simple physical explanation for the intriguing
universality of $\left\langle T_{\mu \nu }\right\rangle$ in the
conformally coupled theory, one may wonder whether the vanishing of
$\Delta \left\langle T_{\mu \nu }\right\rangle $ in the case $\xi
=1/6 $ is actually related to conformal invariance, or whether it is
merely a remarkable numerical coincidence. The question is non
trivial, because the quantity $(\xi -1/6)$ usually arises in
semiclassical computations in dimensions $D \geq 4$, since the
coefficient $a_{1}$ of the Schwinger-De Witt expansion is
$a_1=(\xi-1/6)R$ in all dimensions \cite{BD}. In this
paper we work out a dimensional extension of the computation of \cite%
{Anderson} and show that conformal
invariance is actually playing a crucial role in this phenomenon.

We will perform the explicit computations of the observables $\Delta
\left\langle T_{\mu \nu }\right\rangle $ and $\Delta \left\langle
\phi ^{2}\right\rangle $ in the large distance limit of a
spherically symmetric static space-time in arbitrary number of
dimensions $D$, and with arbitrary coupling $\xi $ between the
scalar field and the curvature. In particular, we will show that the
following expression holds
\begin{equation}
\Delta \left\langle T_{\mu \nu }(x)\right\rangle =\left\langle T_{\mu \nu
}(x)\right\rangle _{\text{Star}}-\left\langle T_{\mu \nu }(x)\right\rangle _{%
\text{BH}}\sim \frac{\xi (\xi -\xi _{D})M}{r^{2D-3}}\text{ }+\mathcal{O}%
(M^{2}/r^{3D-6})  \label{Adelanto}
\end{equation}%
with $\xi _{D}=\frac{(D-2)}{4(D-1)}$, i.e. the conformal coupling in
$D$ dimensions. This implies that the large distance behavior of the
semiclassical correction to the stress tensor of a conformally
invariant scalar field is independent of the nature of the
gravitational source. This manifestly shows that conformal
invariance plays an important role in this universality phenomenon.

An additional motivation to extend the computation of
\cite{Anderson} to higher dimensions would come from the conjectured
correspondence between quantum corrected black holes in
$D$-dimensional braneworlds and classical extended objects in
$D+1$-dimensional bulks \cite{Tanaka,EFK}. Typically, the number of
gravitational solutions with a given asymptotic symmetry is known to
grow as the dimensionality of space-time increases, and, therefore,
it would be natural to ask whether the universality in the
computation of the backreaction effects induced by $\left\langle
T_{\mu \nu }\right\rangle $ is maintained when $D$ becomes larger.
Speculatively, studying the universality of $\left\langle T_{\mu \nu
}\right\rangle $ in the $D$-dimensional conformally coupled theory
might be useful to indirectly learn about the
unicity of extended solutions representing localized objects in $D+1$%
-dimensions. We derive the explicit expression of $\langle T_{\mu
\nu }\rangle $ of a Schwarzschild-Tangherlini black hole in the far
field limit in Section 3.

The explicit computation of $\left\langle T_{\mu \nu }\right\rangle
$ in the $D$-dimensional conformal theory would be also important
within the context of AdS/CFT correspondence \cite{malda}. It is
well known that the so-called Randall-Sundrum Maldacena
complementarity \cite{duff} yields a remarkably numerical agreement
between boundary and bulk computations of the corrected graviton
propagator. This agreement is relatively well understood for $D=4$
where, by means of the
introduction of a IR cut-off, the boundary theory corresponds to the $%
\mathcal{N}=4$ SYM theory coupled to gravity, and non-renormalization
theorems are available. In general, performing such a bulk-boundary
comparison is a highly non-trivial problem, and one has no hope of having an
explicit $D$-dimensional analogue of the computation of \cite{duff}.
Nevertheless, even in this case, having achieved to explicitly compute $%
\left\langle T_{\mu \nu }\right\rangle $ is important, as this
quantity gives the one-loop scalar matter correction to the graviton
propagator in $D$ dimensions \cite{ABF}. This provides important
information about the functional form of both bulk and boundary
quantities.

The paper is organized as follows. In Section 2 we will compute
the differences
 $\Delta \left\langle \phi
^{2}\right\rangle $ and $\Delta \left\langle T_{\mu \nu
}\right\rangle $ for  a massless scalar field in $D$ dimensions,
showing explicitly that both vanish for minimal coupling and that
$\Delta \left\langle T_{\mu \nu }\right\rangle $ vanishes also for
conformal coupling. In Section 3 we compute explicitly  $
\left\langle \phi ^{2}\right\rangle $ and $\left\langle T_{\mu \nu
}\right\rangle $ in the weak field approximation. These results,
combined with the differences computed in Section 2, allow us to
compute the large distance behavior of the vacuum polarization
around a $D$ dimensional Schwarszchild-Tangherlini black hole. As
another application of the results for weak gravitational fields, we
compute the quantum corrections to the Newtonian potential in $D$
dimensions.  Section 4 contains the conclusions of our work.

\section{Universality in the conformally invariant theory}

In this section we will compute the quantities $\Delta \left\langle
\phi ^{2}\right\rangle $ and $\Delta \left\langle T_{\mu \nu
}\right\rangle $, as defined in (\ref{A})-(\ref{B}). This allows to
compare the vacuum polarization effect produced by a star and that
produced by a black hole, both in the large distance limit. First,
we will compute the differentce $\Delta \langle \phi ^{2}\rangle
=\langle \phi ^{2}\rangle _{\text{Star}}-\langle \phi ^{2}\rangle
_{\text{BH}}$ for a massless scalar field in $D$ dimensions and with
arbitrary coupling to the curvature. Then, we will address the
computation of $\Delta \langle T_{\mu \nu }\rangle $ in the large
distance limit. To compute these expectation values we resort to a
dimensional extension of the method developed in \cite{Anderson},
which we will follow closely. Let us briefly review the main steps.

First, consider the Euclidean static spherically symmetric space in $D$
dimensions, with metric%
\begin{equation}
ds^{2}=f(r)d\tau ^{2}+\frac{1}{k(r)}dr^{2}+r^{2}d\Omega _{n}^{2},
\label{metrica}
\end{equation}%
where $f(r)$ and $k(r)$ are two positive functions, and where
$d\Omega _{n}^{2}$ is the line element of the unit $n$-sphere, with
$n=D-2$. In the absence of matter, the metric (\ref{metrica}) is
given by the
Schwarzschild-Tangherlini \cite{tangh} solution\ $f(r)=k(r)=1-\left( \frac{r_{h}}{r}%
\right) ^{n-1}$, and for the black hole case it develops a horizon at $%
r=r_{h}$.

To compute the expectation value $\langle \phi ^{2}\rangle $, let us
be reminded of the fact that the unrenormalized value of $\langle
\phi
^{2}\rangle $ is given by the real part of the Euclidean Green function $%
G_{E}(x,x^{\prime })$ in the coincidence limit $x\rightarrow x^{\prime }$.
Namely
\begin{equation}
\langle \phi ^{2}(x)\rangle =\lim_{x^{\prime }\rightarrow x}\text{Re}
( G_{E}(x,x^{\prime })).
\end{equation}

The differential equation to be obeyed by the Euclidean Green function is
\cite{Anderson}
\begin{equation}
\left( \Box _{x}-\xi R\right) G_{E}(x,x^{\prime })=-\frac{\delta
^{(D)}(x-x^{\prime })}{\sqrt{g}}.  \label{ecmov}
\end{equation}

To solve this equation, it is convenient to consider the form
\begin{equation}
G_{E}(x,x^{\prime })=\frac{1}{\pi }\int_{0}^{\infty }d\omega \cos
\left( \omega \left( \tau -\tau ^{\prime }\right) \right)
\sum_{l,\{m\}}Y_{l\{m\}}^{(n)}(\Omega )Y_{l\{m\}}^{(n)\ast
}(\Omega ^{\prime })R_{l \omega }(r,r^{\prime }),  \label{sep}
\end{equation}%
where $Y_{l\{m\}}^{(n)}(\Omega )$ are the harmonic functions on the $n$%
-sphere, $S^{n}$, satisfying \cite{frolov}
\begin{equation}
\triangle Y_{l\{m\}}^{(n)}(\Omega )=-\frac{l(l+n-1)}{r^{2}}%
Y_{l\{m\}}^{(n)}(\Omega ),
\end{equation}%
being $\Delta $ the Laplacian on $S^{n}$. Then, in the vacuum
region, (\ref{ecmov}) takes the form
\begin{equation}
\partial _{r}^{2}R_{l\omega }(r,r^{\prime })+\left( \frac{n}{r}+(\partial
_{r}\log f)\right) \partial _{r}R_{l\omega }(r,r^{\prime })-\left( \frac{
\omega ^{2}}{f^{2}}+\frac{l(l+n-1)}{fr^{2}}\right) R_{l\omega }(r)=-\frac{
\delta (r-r^{\prime })}{r^{n}},  \label{ecerre}
\end{equation}%
where $f(r)=k(r)=1-\left( \frac{r_{h}}{r}\right) ^{n-1}$.

It is also convenient to factorize $R_{l\omega }(r)$ as follows
\begin{equation}
R_{l\omega }(r,r^{\prime })=C_{\omega l}\text{ }p_{\omega l}(r_{<})q_{\omega
l}(r_{>}),
\end{equation}%
where $r_{>}$ (and $r_{<}$) means the grater (resp. the smaller)
between $r$ and $r^{\prime }$, and where $p_{\omega l}(r)$ and
$q_{\omega l}(r)$ are two independent homogeneous solutions to (\ref{ecerre}).%

In addition, $p_{\omega l}$ and $q_{\omega l}$ satisfy the
Wronskian condition%
\begin{equation}
C_{\omega l}\left( q_{\omega l}^{\prime }(r)p_{\omega
l}(r)-q_{\omega l}(r)p_{\omega l}^{\prime }(r)\right)
=-\frac{1}{f(r)r^{n}},  \label{wro}
\end{equation}%
where the prime denotes the derivative with respect to $r$. This
expression (\ref{wro}) follows from integrating Eq. (\ref{ecerre})
over an infinitesimal region around the point $r'$.

Now, let us compute the quantity $\Delta \langle \phi ^{2}\rangle
\equiv
\langle \phi ^{2}\rangle _{\text{Star}}-\langle \phi ^{2}\rangle _{%
\mathrm{BH}}$. From the expressions above, we can write
\begin{equation}
\Delta \langle \phi ^{2}(x)\rangle =\text{Re}\left( \frac{1}{\pi }
\int_{0}^{\infty }d\omega \sum_{l,\{m\}}Y_{l\{m\}}^{(n)}(\Omega
)Y_{l\{m\}}^{(n)\ast }(\Omega )\left( C_{\omega l}^{\text{Star}}p_{\omega
l}^{\text{Star}}(r)q_{\omega l}^{\text{Star}}(r)-C_{\omega l}^{\mathrm{BH}
}p_{\omega l}^{\mathrm{BH}}(r)q_{\omega l}^{\mathrm{BH}}(r)\right) \right) ,
\label{deltaphi}
\end{equation}
where the superscripts Star and BH label the modes corresponding
to the star and the black hole, respectively. Note that, although
$\langle \phi ^{2}\rangle _{\text{Star}}$ and $\langle \phi ^{2}\rangle _{%
\mathrm{BH}}$ are both divergent quantities, their difference must
be finite outside the star, since the covariant renormalization
involves the subtraction of the Schwinger-DeWitt expansion of the
Green function \cite{BD, christensen}, which is local in the
metric.

The reason why the modes for the star and those for the black hole
differ from each other, is that they must satisfy different
boundary conditions. More precisely, the modes $q_{\omega l}$ must
be regular at infinity, for both star and black hole, so $q_{\omega l}^{%
\text{BH}}=q_{\omega l}^{\text{Star}}=q_{\omega l}$. On the other hand, the
modes $p_{\omega l}^{\text{BH}}$ must be regular at the horizon, while $%
p_{\omega l}^{\text{Star}}$ must be regular at the origin. Such are the
boundary conditions for the two-point function to be well defined in the
region where the Schwarzschild metric holds.

Outside the star, we can write $p_{\omega l}^{\text{Star}}$ as a
linear combination of two independent solutions $p_{\omega
l}^{\mathrm{BH}}$ and $q_{\omega l}$,
\begin{equation}
p_{\omega l}^{\text{Star}}(r)=\alpha _{\omega l}p_{\omega l}^{\mathrm{BH}%
}(r)+\beta _{\omega l}q_{\omega l}(r).  \label{comb}
\end{equation}

In turn, coefficients $\beta _{\omega l}$ mix the modes in the star
background. The reader may refer to Ref. \cite{Anderson} for further details.

By evaluating Eq. (\ref{wro}) for both the case of the star and the case of
the black hole, and using (\ref{comb}), we get the relation $\alpha _{\omega
l}C_{\omega l}^{\text{Star}}=C_{\omega l}^{\mathrm{BH}}$, so we get
\begin{equation}
\Delta \langle \phi ^{2}(x)\rangle =\text{Re}\left( \frac{1}{\pi }%
\int_{0}^{\infty }d\omega \sum_{l,\{m\}}Y_{l\{m\}}^{(n)}(\Omega
)Y_{l\{m\}}^{(n)\ast }(\Omega )C_{\omega l}^{\mathrm{BH}}\frac{\beta
_{\omega l}}{\alpha _{\omega l}}(q_{\omega l})^{2}\right) .
\label{deltaphi2}
\end{equation}

As we are interested in the region far from the gravitational bodies, we
consider the leading contribution in the $1/r$ expansion. Consequently, we
are interested in the flat space modes

\label{modflat}
\begin{equation}
q_{\omega l}^{\mathrm{flat}}(r)=r^{1-\frac{n}{2}}\omega
^{a+1}k_{a}(\omega r),\qquad p_{\omega
l}^{\mathrm{flat}}(r)=r^{1-\frac{n}{2}}\omega ^{-a}i_{a}(\omega r),
\end{equation}%
where $k_{a}$ and $i_{a}$ are the modified spherical Bessel
functions with $a=l+ \frac{n}{2}-1$. In turn, the Wronskian
condition reads $C_{\omega l}^{\mathrm{BH}}=\frac{2}{\pi }$.

Dimensional analysis, combined with the mean value theorem, leads to
the conclusion that only the $\omega =l=0$ contribution is relevant
in the $1/r$ expansion, yielding the result
\begin{equation}
\Delta \langle \phi ^{2}(x)\rangle =\frac{(n-1)}{16\pi ^{\frac{n+1}{2}
}r^{2n-1}}\Gamma \left( \frac{n-1}{2}\right) \text{Re}\left( \frac{\beta
_{\omega =0,l=0}}{\alpha _{\omega =0,l=0}}\right) \frac{\Gamma \left( \frac{n
}{2}\right) \Gamma \left( n-\frac{1}{2}\right) }{\Gamma \left( \frac{n+1}{2}
\right) }.  \label{deltaphi7}
\end{equation}

This is valid for any static spherically symmetric star. It is worth
noticing that this quantity vanishes for $\xi =0$. This is because,
when $\omega =l$ $=0$ and $\xi =0$, the homogeneous solutions to
(\ref{ecerre}) that have to be regular at the black hole horizon, or
regular at the center of the star, are constant. Then, because of
the relation (\ref {comb}) and because $q_{\omega =0,l=0}$ is not a
constant, $\beta _{\omega =0,l=0}$ must be zero. Actually, it would
be convenient to keep in mind that $\beta _{\omega =0,l=0}$\ is
proportional to $\xi $.

The result for  $\Delta \langle \phi ^{2}\rangle $ depends on the
inner structure of the star through the factor $\text{Re}\left(
\frac{\beta _{\omega =0,l=0}}{\alpha _{\omega =0,l=0}}\right)$. Now,
let us compute this factor explicitly for the case of a
weakly gravitating star. First, we can perturbe the modes as follows%
\begin{equation}
p_{\omega =0,l=0}(r)=p_{\omega =0,l=0}^{\mathrm{flat}}(r)+\delta
p(r)\,,\,\,q_{\omega =0,l=0}(r)=q_{\omega
=0,l=0}^{\mathrm{flat}}(r)+\delta q(r)
\end{equation}%
being $\delta p$ and $\delta q$ small perturbations around flat
solutions

\begin{equation}
p_{\omega =0,l=0}^{\mathrm{flat}}=\frac{\sqrt{\pi }2^{-\frac{n}{2}}}{\Gamma
\left( \frac{n+1}{2}\right) },\qquad q_{\omega =0,l=0}^{\mathrm{flat}}(r)=\frac{%
\sqrt{\pi }2^{\frac{n}{2}-2}}{r^{n-1}}\Gamma \left( \frac{n-1}{2}\right) .
\label{modceroflat}
\end{equation}

By writing $\alpha _{\omega l}$ and $\beta _{\omega l}$ in terms of the
modes and their first derivatives, and keeping only first order terms, one
gets

\begin{equation}
\frac{\beta _{\omega =0,l=0}}{\alpha _{\omega =0,l=0}}=\frac{(\delta p^{%
\text{Star}\prime }-\delta p^{\text{BH}\prime })}{q_{\omega =0,l=0}^{\mathrm{%
flat}\prime }}\Big|_{r=r^{\ast} } \label{abdesa}
\end{equation}

Where, again, the prime means the derivative with respect to $r$.
Then, it remains to compute $\delta p^{\text{Star}\prime }$ and $\delta p^{%
\text{BH}\prime }$ evaluated at the radius of the star $r^{\ast }$. The
latter is exactly zero, as it turns out that $p_{\omega =0,l=0}^{\text{BH}%
}=p_{\omega =0,l=0}^{\text{flat}}$. On the other hand, by solving the
linearized differential equation for $\delta p^{\text{Star}}$, and demanding
regular behavior at the origin, we find

\begin{equation}
\frac{d}{dr}\delta p^{\text{Star}}(r) =\xi \frac{\sqrt{\pi }2^{-\frac{n}{2}}}{%
\Gamma \left( \frac{n+1}{2}\right) }\frac{1}{r^{n}}\int_{0}^{r}dr\,r^{n}R(r).
\label{deltapes}
\end{equation}

Now, it is possible to evaluate expression (\ref{abdesa}) as a
function of $D $. Using (\ref{deltaphi7}), we eventually find
\begin{equation}
\Delta \langle \phi ^{2}(x)\rangle =-\xi 2^{3-D}\pi ^{2-D}\frac{M}{r^{2D-5}}%
\frac{\Gamma \left( \frac{D}{2}-1\right) \Gamma \left( D-\frac{5}{2}\right)
}{(D-2)\Gamma \left( \frac{D-1}{2}\right) }.  \label{deltaphifinal}
\end{equation}

Here we additionally used the identity $\int d^{D-1}x
R(x)=\frac{16\pi M}{D-2}$ which holds for any static mass
distribution. This allows us to claim that (\ref{deltaphifinal}) is
independent of the internal structure of the weakly gravitating
star. Expression (\ref{deltaphifinal}) is the difference between
$\langle \phi ^{2}\rangle $ computed for a weakly gravitating star
and the same quantity computed for a black hole of the same mass in
the region far from these objects. It is worth mentioning that this
result agrees with that of \cite{Anderson} for the case $D=4$.

Now, we move on to compute the quantity $\Delta \langle T_{\mu \nu }\rangle
=\langle T_{\mu \nu }\rangle _{\text{Star}}-\langle T_{\mu \nu }\rangle _{%
\text{BH}}$, which corresponds to the far field limit of the
difference between the expectation value $\langle T_{\mu \nu
}\rangle $ for a static spherically symmetric star and that for a
Schwarzschild-Tangherlini black hole. Since the computation of
$\Delta \langle T_{\mu \nu }\rangle $ is quite similar to that of
$\Delta \langle \phi ^{2}\rangle $ we discussed above, and in order
to avoid redundancies, we will limit ourself to present the results.
The reader can find the details in \cite{Anderson}.

To compute $\langle T_{{\mu \nu }}\rangle $, it is convenient to
write this quantity as the coincidence limit of the Euclidean
Green function $G_{E}(x,x^{\prime })$ and of its covariant
derivatives $G_{E;\mu ^{\prime }\nu }=\nabla _{\mu ^{\prime
}}\nabla _{\nu }G_{E}(x,x^{\prime })$. Namely \cite{ASH},
\begin{eqnarray}
\langle T_{{\mu \nu }}(x)\rangle  &=&\lim_{x^{\prime }\rightarrow
x}\left( \left( \frac{1}{2}-\xi \right) \left( g_{\mu }^{\alpha
^{\prime }}G_{E;\alpha ^{\prime }\nu }+g_{\nu }^{\alpha ^{\prime
}}G_{E;\alpha ^{\prime }\mu }\right) +\left( 2\xi
-\frac{1}{2}\right) g_{{\mu \nu } }g^{\alpha ^{\prime }\sigma
}G_{E;\alpha ^{\prime }\sigma }\right.   \notag
\label{tmngreen} \\
&&\left. -\xi \left( G_{E;{\mu \nu }}+g_{\mu }^{\alpha ^{\prime }}g_{\nu
}^{\beta ^{\prime }}G_{E;\alpha ^{\prime }\beta ^{\prime }}\right) \right) .
\end{eqnarray}

Then, following similar steps to those described above, and after some lengthy
calculations, we find the following results for the differences $\Delta
\langle T_{\nu }^{\mu }\rangle $,

\begin{equation}
\Delta \langle T_{\mu }^{\nu }(x)\rangle =\frac{(D-2)2^{6-3D}\pi ^{-\frac{D-1}{2%
}}}{r^{2D-3}}\frac{\Gamma (2D-4)}{\Gamma
(\frac{D-1}{2})}\text{Re}\left(\frac{\beta
_{\omega =0,l=0}}{\alpha _{\omega =0,l=0}}\right)(\xi -\xi _{D})\,\mathrm{diag}%
\left( 1,-1,\frac{D-1}{D-2},...,\frac{D-1}{D-2}\right) .  \label{DeltaTuv}
\end{equation}%
As in the case of $\Delta \langle \phi ^{2}\rangle $, this quantity
is found to vanish in the minimally coupled theory, because $\beta
_{\omega =0, l=0 }$ is proportional to $\xi$. Then, replacing in
(\ref{DeltaTuv}) the value of $\frac{\beta _{\omega =0, l=0
}}{\alpha_{\omega=0, l=0}}$ that corresponds to a weakly gravitating
star, we find
\begin{equation}
\Delta \langle T_{\mu }^{\nu }(x)\rangle =-\xi (\xi -\xi _{D})\frac{%
2^{12-4D}\pi ^{3-D}M}{r^{2D-3}}\frac{\Gamma (2D-4)}{\Gamma (\frac{D-1}{2}%
)^{2}}\times \mathrm{diag}\left( 1,-1,\frac{D-1}{D-2},...,\frac{D-1}{D-2}
\right) .  \label{DeltaTuvdebil2}
\end{equation}
As expected, this expression agrees with that of \cite{Anderson}
in the particular case $D=4$.

\section{Expectation values on a weakly gravitating background}

In this section we will make use of the results of \cite{Avra} to
calculate the expectation values $\langle \phi ^{2}\rangle $ and
$\langle T_{\mn}\rangle $ for a weakly gravitating object.

Let us start by considering equation $(23)$ in \cite{Avra}, from which we can write the one-loop effective action $%
\Gamma _{(1)}$ for $D>2$ as follows
\begin{eqnarray}
\Gamma _{(1)} &=&\frac{1}{2}(4\pi )^{-D/2}\int d^{D}x\sqrt{g}\left( \xi
^{2}R\beta _{D/2}^{(1)}(\Box )R-2\xi R\beta _{D/2}^{(3)}(\Box )R+R_{{\mu \nu
}}\beta _{D/2}^{(4)}(\Box )R^{{\mu \nu }}\right.   \notag \\
&&\left. +R\beta _{D/2}^{(5)}(\Box )R+\mathcal{O}(R^{3})\right) ,
\label{AVVV}
\end{eqnarray}%
where the functions $\beta _{D/2}^{(i)}(\Box )$ are given by%
\begin{equation*}
\beta _{D/2}^{(i)}(\Box )=\frac{\sqrt{\pi }}{4}\frac{(-1)^{D/2}}{\Gamma
((D-1)/2)}f_{D/2}^{(i)}\left( -\frac{\Box }{4}\right) ^{D/2-2}\ln \frac{%
-\Box }{\mu ^{2}}
\end{equation*}%
for even $D$, while
\begin{equation}
\beta _{D/2}^{(i)}(\Box )=\frac{1}{4}\pi ^{3/2}\frac{(-1)^{(D-1)/2}}{\Gamma
((D-1)/2)}f_{D/2}^{(i)}\left( -\frac{\Box }{4}\right) ^{D/2-2}  \label{beta}
\end{equation}%
for odd $D$. The factors $f_{D/2}^{(i)}$ in (\ref{AVVV}) are given by
\begin{eqnarray}
f_{D/2}^{(1)} &=&1,\quad \quad \qquad \qquad \qquad
f_{D/2}^{(3)}=\xi _{D}=\frac {D-2}{4(D-1)},
\notag  \label{efes} \\
f_{D/2}^{(4)} &=&\frac{1}{2(D-1)(D+1)},\quad f_{D/2}^{(5)}=\frac{%
(D/2)^{2}-D/2-1}{4(D-1)(D+1)}.
\end{eqnarray}

In order to compute $\langle \phi ^{2}\rangle $ in $D$ dimensions,
one could address the calculation by using  a resumation of the
Schwinger-DeWitt expansion \cite{vilko}, or by computing
perturbatively the two point function, along the lines of Ref.
\cite{satz}. However, even when these methods lead to the right
expression, here we prefer to take a shortcut by exploiting the fact
that varying the effective action with respect to $\xi $ yields

\begin{equation}
\frac{d}{d\xi }e^{-\Gamma _{(1)}}=\int \left[ \mathcal{D}\phi
\right] \,\,\frac{d}{d\xi}e^{-S[g_{\mn},\phi ]}=\frac{1}{2}\int
d^{D}x\,\sqrt{g}R\langle \phi ^{2}\rangle ,
\end{equation}%
so that one can read $\langle \phi ^{2}\rangle $ form this expression directly.
Varying (\ref{AVVV}) with respect to $\xi $ and then performing a Wick
rotation, we find
\begin{equation}
\langle \phi ^{2}(x)\rangle =-\frac{-\pi ^{\frac{D+1}{2}}}{(2\pi )^{D}}(-1)^{%
\frac{D}{2}}\frac{2^{3-D}}{\Gamma (\frac{D-1}{2})}\,(\xi -\xi _{D})(-\Box )^{%
\frac{D}{2}-2}\ln \frac{-\Box }{\mu ^{2}}R(x).  \label{phi2par}
\end{equation}%
On the other hand, the analogous expression for odd dimensions reads
\begin{equation}
\langle \phi ^{2}(x)\rangle =\frac{-\pi ^{\frac{D+3}{2}}}{(2\pi )^{D}}(-1)^{%
\frac{D+1}{2}}\frac{2^{3-D}}{\Gamma (\frac{D-1}{2})}\,(\xi -\xi _{D})(-\Box
)^{\frac{D}{2}-2}R(x).  \label{phi2impar}
\end{equation}

It is worth noticing that for a weakly gravitating object the
expectation value $\langle \phi ^{2}\rangle$ vanishes in the
conformally coupled case. The reason is the following: Being a
scalar, on general grounds we expect
\begin{equation}
\langle \phi ^{2}(x)\rangle = \left (F_1(-\Box) + \xi F_2(-\Box)\right )R
\end{equation}
for adequate form factors $F_i(-\Box)$.  As this equation must be
valid for any metric, we can specialize it for a metric which is
conformally flat and  asymptotically flat in the past. In this
situation, it is clear that  $\langle \phi ^{2}(x)\rangle$ must
vanish for conformal coupling, since the conformal vacuum coincides
with the IN vacuum. Therefore we conclude that $F_1(-\Box)=-\xi_D
F_2(-\Box)$, i.e. $\langle \phi ^{2}(x)\rangle$ is proportional to
$(\xi -\xi_D)$.

From  expressions  (\ref{phi2par}) and (\ref{phi2impar})  we can obtain the explicit form of
$\langle \phi ^{2}\rangle $ for a static spherically symmetric
star in the far field limit. So, imposing these conditions we get

\begin{equation}
\langle \phi ^{2}(x)\rangle _{\text{Star}}=-(\xi -\xi _{D})2^{3-D}\pi ^{2-D}%
\frac{M}{r^{2D-5}}\frac{\Gamma \left( \frac{D}{2}-1\right) \Gamma \left( D-%
\frac{5}{2}\right) }{(D-2)\Gamma \left( \frac{D-1}{2}\right) },
\label{phistar}
\end{equation}%
which is valid for arbitrary number of dimensions $D>2$.

As a simple consistency check of the calculation above we can
compare the term that is linear in $\xi $ in both $\langle \phi
^{2}\rangle _{\text{Star}}$ and $\Delta \langle \phi ^{2}\rangle $.
Since no dependence on $\xi $ appears in the mode equation for
$p_{\omega ,l}^{\text{BH}}$ and $q_{\omega ,l}^{\text{BH}}$, then
the quantity $\langle \phi ^{2}\rangle _{\text{BH}}$ turns out to be
independent of that coupling constant. In other words, we verify
$\Delta \langle \phi
^{2}\rangle _{|_{\mathcal{O}(\xi )}}=\langle \phi ^{2}\rangle _{%
\mathrm{Star}|_{\mathcal{O}(\xi )}}$.

Notice also that expression (\ref{phistar}) permits to obtain
$\langle \phi ^{2}\rangle $ in the black hole background in the
region far from the horizon. In fact, using (\ref{deltaphifinal}) we
find that in $D$ dimensions this quantity is given by
\begin{equation}
\langle \phi ^{2}(x)\rangle _{\mathrm{BH}}=\xi _{D}2^{3-D}\pi ^{2-D}\frac{M}{%
r^{2D-5}}\frac{\Gamma \left( \frac{D}{2}-1\right) \Gamma \left( D-\frac{5}{2}%
\right) }{(D-2)\Gamma \left( \frac{D-1}{2}\right) }. \label{phiBH}
\end{equation}

On the other hand, the expectation value $\langle T^{{\mu \nu
}}\rangle $ in a weak field background is given by varying the
effective action with respect to the metric, and writing the
result up to second order in the curvature; namely
\begin{equation}
\langle T^{{\mu \nu }}(x)\rangle=-\frac{2}{\sqrt{g}}\frac{\delta \Gamma _{(1)}%
}{\delta g_{{\mu \nu }}}+\mathcal{O}(R^{2}),  \label{tmneuc}
\end{equation}

This expression can be written down in the following way
\begin{equation}
\langle T_{{\mu \nu }}(x)\rangle =(\xi -\xi _{D})^{2}A_{{\mu \nu
}}+B_{{\mu \nu }},  \label{Tuvesquema}
\end{equation}%
where
\begin{eqnarray}
A_{{\mu \nu }} &=&f_{D}F(\Box )H_{{\mu \nu }}^{(1)},  \label{defBmn} \\
B_{{\mu \nu }} &=&f_{D}\left( (f_{D/2}^{(5)}-\xi _{D}^{2})F(\Box )H_{{\mu
\nu }}^{(1)}+f_{D/2}^{(4)}F(\Box )H_{{\mu \nu }}^{(2)}\right)
\end{eqnarray}%
and, for even dimensions,
\begin{eqnarray}  \label{fd}
F(\Box)&=&(-\Box)^{\frac{D}{2}-2}\ln \frac{-\Box}{\mu^2} \\
f_D &=& \frac{\pi^{3D/2} 2^{D/2} (-1)^{D/2+1}}{(2\pi)^{2D}(D-3)!!} \\
H_\mn^{(1)} &=& 4\nabla_\mu\nabla_\nu R-4g_\mn\Box R+\mathcal{O}(R^2) \\
H_\mn^{(2)} &=& 2\nabla_\mu\nabla_\nu R-g_\mn\Box R-2\Box R_\mn+\mathcal{O}%
(R^2), \label{ds}
\end{eqnarray}

The term $B_{\mu \nu }$ in (\ref{Tuvesquema}) is the only one that
contributes in the conformal invariant case $\xi =\xi _{D}$. Such
contribution can be seen to be traceless, so it does not appear in
the trace anomaly, and $\langle T_{\mu }^{\mu }\rangle $ vanishes.
This is because the anomaly is of higher order in the curvature.

The case of odd dimension $D$ is similar. In fact, it follows from (\ref{Tuvesquema})-(\ref{ds})
by replacing $f_{D}$ and $F(\Box )$ in the expressions above by%
\begin{eqnarray}
\tilde{f}_{D} &=&2^{1-D}\pi ^{3/2}(4\pi )^{-D/2}\frac{(-1)^{\frac{D+1}{2}}}{%
\Gamma (\frac{D-1}{2})} \\
\tilde{F}(\Box ) &=&(-\Box )^{D/2-2},
\end{eqnarray}
which come from (\ref{beta}).

Once spherical symmetry and staticity are imposed, expression (\ref%
{Tuvesquema}) yields the following result for the expectation value
of the stress tensor in the region far away from the star,
\begin{eqnarray}
\langle T_{\mu }^{\nu }(x)\rangle _{\text{Star}}
&=&-\frac{2^{10-4D}\pi ^{3-D}M }{r^{2D-3}}\frac{\Gamma
(2D-4)}{\Gamma (\frac{D-1}{2})^{2}}\left( 4\left( (\xi -\xi
_{D})^{2}+f_{D/2}^{(5)}-\xi _{D}^{2}\right) \right.   \notag
\label{Tmnestrella} \\
&&\left. \times \mathrm{diag}\left( 1,-1,\frac{D-1}{D-2},...,\frac{D-1}{D-2}%
\right) \right.   \notag \\
&&\left. +f_{D/2}^{(4)}\,\mathrm{diag}\left( 4-D,-2,\frac{2(D-1)}{D-2},...,%
\frac{2(D-1)}{D-2}\right) \right) .
\end{eqnarray}
which is valid in arbitrary number of dimensions $D\geq 4$.

Now, from (\ref{DeltaTuvdebil2}) and (\ref{Tuvesquema}) we can write
$\tmn$ for the case of a Schwarzschild-Tangherlini black hole
background in the region far from the horizon; namely
\begin{eqnarray}
\langle T_{\mu }^{\nu }(x)\rangle _{\mathrm{BH}} &=&-\frac{2^{10-4D}\pi ^{3-D}M%
}{r^{2D-3}}\frac{\Gamma (2D-4)}{\Gamma (\frac{D-1}{2})^{2}}\left( 4\left(
\xi _{D}(\xi _{D}-\xi )+f_{D/2}^{(5)}-\xi _{D}^{2}\right) \right.  \notag
\label{TmnBH} \\
&&\left. \times \mathrm{diag}\left( 1,-1,\frac{D-1}{D-2},...,\frac{D-1}{D-2}%
\right) \right.  \notag \\
&&\left. +f_{D/2}^{(4)}\,\mathrm{diag}\left( 4-D,-2,\frac{2(D-1)}{D-2},...,%
\frac{2(D-1)}{D-2}\right) \right) .  \label{esto}
\end{eqnarray}
It is important to emphasize that this last result, together with
$\langle \phi^2\rangle$ (see (\ref{phiBH})), are vacuum expectation
values for the black hole background in the far field limit computed
entirely with analytical methods, i.e. without the aid of numerical
computations.

As an application of (\ref{Tmnestrella}) we can address the
calculation of the semiclassical correction to the Newtonian
gravitational potential \cite{dalvit}. To do this, we write the
semiclassical Einstein equations using $\langle T_{\mu \nu }\rangle
_{\mathrm{Star}}$ as a source. In the Lorentz gauge, the quantum
corrections to the metric satisfy
\begin{equation}  \label{ech}
\Box h_\mn(x)=-16\pi\left(\langle T_\mn(x)\rangle_{\text{Star}}+\frac{\eta_\mn}{D-2%
}\langle T_\lambda^\lambda(x)\rangle_{\text{Star}}\right).
\end{equation}
Then, by making use of (\ref{Tmnestrella}), we get
\begin{equation}  \label{correccion}
\Phi(r)=-\frac{2^{15-4D} \pi^{4-D} \Gamma(2D-5)}{(D-2)^2\Gamma(\frac{D-1}{2}%
)^2}\frac{M}{r^{2D-5}}\left( (\xi-\xi_D)^2+\frac{(D-2)^3}{8(D-1)^2(D+1)}%
\right) .
\end{equation}

It is worth pointing out that this expression, in the special case
$D=4$ and $\xi=1/6$, agrees with the semiclassical correction to the
gravitational potential \cite{satz,duff}, namely
\begin{equation}
V(r)=-\frac{MG}{r}\left(1+\frac{1}{45\pi}\frac{G}{r^2}\right),
\end{equation}
where we have reintroduced the four dimensional Newton constant $G$
for major clarity. This also agrees with the one-loop correction to
the graviton propagator in the conformally coupled theory
\cite{ABF,capper,duff74}.

\section{Discussion}

Motivated by the question about the connection between conformal
invariance and the universality phenomenon discussed in \cite{Anderson}%
, we addressed the explicit computation of the observables $\Delta
\langle T_{\mu }^{\nu }\rangle $ and $\Delta \langle \phi
^{2}\rangle $, defined as in (\ref{A})-(\ref{B}), in an arbitrary
number of dimensions. These observables gather the vacuum
polarization effects for the case of a massless scalar field in a
static spherically symmetric background. We have shown that in the
$D$-dimensional theory both observables vanish for minimal coupling,
and that $\Delta \langle T_{\mu }^{\nu }\rangle $ also vanishes in
the  conformally coupled theory. This result extends the results of
\cite{Anderson} to $D \geq 4$ dimensions.

Then, using the one-loop effective action, we computed $ \langle
T_{\mu}^{\nu}\rangle $ for a weakly gravitating object. This,
together with the expression for $\Delta \langle T_{\mu }^{\nu
}\rangle $, enabled us to write down the explicit expression of the
expectation value $\langle T_{\mu}^{\nu}\rangle $ for a
Schwarzschild-Tangherlini black hole. As an application of our
results, we obtained the quantum correction to the gravitational
potential in $D$ dimensions, which for $D=4$ are seen to agree with
the one-loop correction to the graviton propagator previously found
in the literature. It is worth mentioning that the functional form
of the quantum correction to the $D$-dimensional gravitational
potential we obtained, agrees with the classical correction induced
by an extra dimension in the Randall-Sundrum scenario \cite{RS,GKR},
both yielding a $1/r^{2D-5}$ dependence in the (corrected) Newtonian
potential. This is to be expected, as the classical action in this
scenario reproduces the nonlocal effective action  given in
(\ref{AVVV}) when restricted to the brane \cite{Alvarez}.

Even though the explicit computation we carried out in Section 2 can
be regarded as a proof of the vanishing of $\Delta \langle T_{\mu
}^{\nu }\rangle $ in both the minimally and conformally coupled
theory, one might still wonder whether an intuitive physical
explanation for this phenomenon exists. Actually, there is a
particular case in which the universality can be demonstrated using
simple arguments. Let us consider a massless field in $D=2$, where
$\xi=0$ corresponds both for minimal and conformal coupling. For a
two-dimensional metric of the form
\begin{equation}
ds^{2}=f(r)d\tau ^{2}+\frac{1}{k(r)}dr^{2},
\label{metrica2d}
\end{equation}
it is well known \cite{christfull} that the conservation law $\nabla
_{\nu }\langle T_{\mu}^{\nu}\rangle=0$ together with the trace
anomaly determine the expectation value $\langle
T_{\mu}^{\nu}\rangle$ (in particular for the Boulware state, when
chosen the appropriate boundary conditions). Therefore, since the
trace anomaly $\langle T^\mu_\mu\rangle=R/24\pi $ depends locally on
the metric, one can show that all the components of $\langle
T_{\mu}^{\nu}\rangle$ are determined by the local values of $f(r)$
and $k(r)$. Probably, a similar intuitive explanation for the
universality in $D>2$ dimensions could be found by analyzing the
dimensionally reduced two-dimensional theory that describes the
$s$-wave sector of the quantum scalar field. However, in absence of
such an intuitive explanation, and given the fact that in $D>2$
dimensions the components of $\langle T_{\mu}^{\nu}\rangle$ are not
fully determined by the trace anomaly, one has to resort to the
computations of Section 2 to explain the so called apparent
universality.

\section*{Acknowledgements}
This work was supported by Universidad de Buenos Aires, ANPCyT,
and CONICET. Conversations with P. Anderson are acknowledged. A.G.
and G.G. thank the members of the Centro de Estudios
Cient\'{\i}ficos CECS for their hospitality. G.G. also thanks the
hospitality of the members of CCPP during his stay at New York University.

\end{document}